\documentclass[english,titlepage,reqno]{amsart}

\usepackage[T1]{fontenc}
\usepackage[latin9]{inputenc}
\usepackage[a4paper]{geometry}
\usepackage{layout}
\usepackage{array}
\usepackage{amsmath}
\usepackage{amssymb}
\usepackage{graphicx}
\usepackage{cancel}

\makeatletter

\providecommand{\tabularnewline}{\\}

\usepackage{youngtab}
\usepackage{cite}
\usepackage[usenames,dvipsnames]{pstricks} 
\usepackage{epsfig} 
\usepackage{pst-grad} % For gradients 
\usepackage{pst-plot} % For axes 
\usepackage[space]{grffile} % For spaces in paths 
\usepackage{etoolbox} % For spaces in paths 
\makeatletter % For spaces in paths 
\patchcmd\Gread@eps{\@inputcheck#1 }{\@inputcheck"#1"\relax}{}{} 
\makeatother % % User Packages: 
\newcommand{\fund}{\tiny\Yvcentermath1\yng(1)}
\newcommand{\afund}{\tilde{\tiny\Yvcentermath1\yng(1)}}

\newcommand{\diag}{\mbox{\rm \tiny{diag}}}

\definecolor{rossoCP3}{cmyk}{0,.88,.77,.40}

\makeatother

\usepackage{babel}
\usepackage{amsaddr}

\begin{document}

\title{A complete asymptotically safe embedding of the Standard Model}

\author{Steven~Abel$^{1,*}$,  Esben~M\o lgaard$^{2,\dagger}$, Francesco~Sannino$^{2,3,\ddag}$}

\address{$^1$IPPP, Durham University, South Road, Durham, DH1 3LE}
\address{$^2${\color{rossoCP3}CP${}^3$-Origins} \& the Danish Institute for Advanced Study,
Univ. of Southern Denmark, Campusvej 55, DK-5230 Odense }
\address{$^3$SLAC, National Accelerator Laboratory, Stanford University, Stanford, CA 94025.}

\email{$^*$s.a.abel@durham.ac.uk}
\email{$^\dagger$molgaard@cp3-origins.net}
\email{$^\ddag$sannino@cp3-origins.net}

\begin{abstract}
We present and discuss the ``Tetrad Model'', a large colour/flavour embedding of the Standard model which has an interacting ultraviolet fixed point. 
It is shown that its extended-Pati-Salam symmetry is broken radiatively via the Coleman-Weinberg mechanism, while the remaining electroweak 
symmetry is broken when mass-squared terms run negative.  In the IR the theory yields just the Standard Model, augmented by the 
fact that the Higgs fields carry the same generation indices as the matter fields.  
It is also shown that the Higgs mass-squareds develop a hierarchical structure in the IR, from a UV theory that is asymptotically flavour symmetric, opening up an 
interesting direction for explaining the emergence of the observed  flavour structure. 
\vskip10.7cm

{\flushleft \noindent \footnotesize Preprint: CP3-Origins-2018-048 DNRF90, 
IPPP/18/106}
\end{abstract}

\maketitle

\section{Introduction and overview of embedding}

A recent series of papers \cite{Abel:2017ujy,Abel:2017rwl} suggested
a framework for embedding the Standard Model (SM) in an asymptotically
safe ultra-violet (UV) completion \cite{weinberg}. (For some earlier discussions of asymptotic safety applications  
 see \cite{Martin:2000cr,Gies:2003dp,shapo,Gies:2009sv,Braun:2010tt,Bazzocchi:2011vr,Wetterich:2011aa,chacko,Antipin:2013pya,Gies:2013pma,late7,Abel:2013mya,Litim:2014uca,Litim:2015iea,Intriligator:2015xxa,Bond:2016dvk,Pelaggi:2017abg,Bond:2017wut}. For recent reviews see \cite{Litim:2011cp,Eichhorn:2018yfc}.) The framework
is partially perturbative based on the weak ultra-violet (UV) fixed
points of \cite{Antipin:2013pya,Litim:2014uca,Litim:2015iea} (hereafter LS). These are the UV counterparts
of the better known Caswell-Banks-Zaks infra-red (IR) fixed points,
in a large colour and flavour Veneziano limit. By suitable adjustment of the 
numbers of colours and flavours, a UV fixed point can be achieved that is 
arbitrarily weakly coupled. 
Coupled with the ``large
flavour'' fixed points of \cite{gracey,Pica:2010xq,Molinaro:2018kjz,Antipin:2018zdg,Antipin:2017ebo,Mann:2017wzh} operating for the electroweak gauge couplings, 
one finds an asymptotically
safe extension of the Pati-Salam (PS) theory, that has a UV fixed point
with gauge group $SU(N_{C})\times SU(2)_{L}\times SU(2)_{R}$,
and a natural breaking down to the SM gauge group in the IR driven
partially by radiative symmetry breaking. The main observation of
\cite{Abel:2017rwl} was that the two kinds of fixed points (Veneziano
and large $N_{f}$) do not interfere with each other. 

Despite this attractive framework for embedding the Standard Model, the 
theories presented in  \cite{Abel:2017ujy,Abel:2017rwl} did not provide a mechanism for 
fully removing the 
extraneous degrees of freedom in the IR to leave {\it purely} the SM. 
In particular in this simplest realisation, there remain in the low energy theory a large multiplicity of 
electroweak $SU(2)$ doublets, that are unmatched and hence massless.  

In this paper we provide a complete phenomenological framework, by an enhancement
that yields a theory flowing from an asymptotically safe fixed point in the UV to precisely
the SM, augmented only by additional Higgses. In particular there are no other light
superfluous states remaining in the theory. The additional Higgses are furnished with the 
same generation numbers as the matter fields, so their VEVs 
may therefore ultimately be able to explain flavour hierarchies (although we do not attempt this in the 
present paper). 

Moreover symmetry breaking can be
entirely radiative. It can happen in two ways (or by a combination of them). One possibility is the 
traditional radiative symmetry breaking mechanism of Coleman and Weinberg  \cite{Coleman:1973jx,gildener2,gildener}. This can be 
shown to occur analytically driven by a single quartic coupling running negative and generating a minimum  according to the 
pattern discussed in \cite{gildener2}. This can be responsible for the bulk of the breaking of the extended PS gauge group.
At the same time the PS breaking generates a positive mass-squared for the Higgs at the high scale due to a portal-like coupling between the 
electroweak Higgs and the PS Higgs. This can run negative  in the IR due to large Yukawa couplings from its initially positive boundary value at the PS-scale. 
The alternative possibility is that radiative symmetry breaking is instead dominated by the dimensionful  couplings (i.e. mass-squared terms) and the 
PS breaking minimum is generated radiatively when they run  
negative due to the large Yukawa couplings that have to be present in the LS gauge-Yukawa theories. This was the 
mechanism discussed in \cite{Abel:2017ujy}, which is essentially the asymptotically safe version of the radiative symmetry breaking in the 
supersymmetric Standard Model \cite{Ibanez:1982fr}. 
Thus one appears to have the freedom to turn on as much or as little of the classically dimensionful operators as desired in the symmetry breaking.

In order to present our model we will also advocate in this paper the use of  ``quiver'' diagrams. 
Such diagrams can greatly alleviate the generic problem that the UV of asymptotically safe models is 
 complicated because they necessarily have to include extra degrees of freedom, 
 and as a consequence the structure is often hard to appreciate (or present), even though it may 
in reality  be relatively simple. Although 
(depending on the model in question) it may only be possible to represent part of the gauge groups in quiver diagrams
(if some of the states do not easily fall into bi-fundamentals), their use can greatly ease the construction 
of phenomenologies within an asymptotically safe framework.      
 
\section{The ``Tetrad'' Model (TM)}

\subsection{\it Structure in the UV}
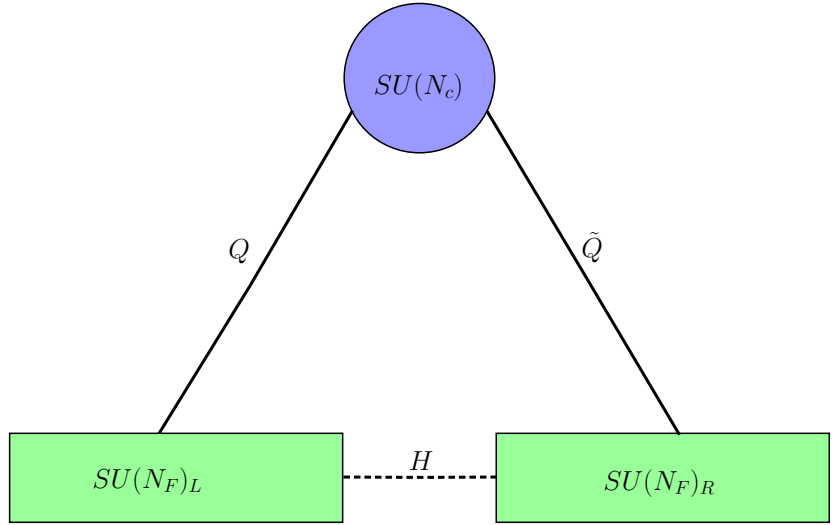
\begin{figure}

% \usepackage[usenames,dvipsnames]{pstricks} 
% \usepackage{epsfig} 
% \usepackage{pst-grad} % For gradients 
% \usepackage{pst-plot} % For axes 
% \usepackage[space]{grffile} % For spaces in paths 
% \usepackage{etoolbox} % For spaces in paths 
% \makeatletter % For spaces in paths 
% \patchcmd\Gread@eps{\@inputcheck#1 }{\@inputcheck"#1"\relax}{}{} 
% \makeatother % % User Packages: 
%  % 
\begin{center}
\psscalebox{0.5 0.5} % Change this value to rescale the drawing. 
{ \begin{pspicture}(0,-6.9)(21.6,6.9) \definecolor{colour0}{rgb}{0.6,1.0,0.6} \definecolor{colour1}{rgb}{0.6,0.6,1.0} \psframe[linecolor=black, linewidth=0.04, fillstyle=gradient, gradlines=2000, gradbegin=colour0, gradend=colour0, dimen=outer](21.6,-4.5)(12.8,-6.9) \pscircle[linecolor=black, linewidth=0.04, fillstyle=gradient, gradlines=2000, gradbegin=colour1, gradend=colour1, dimen=outer](10.8,4.9){2.0} \psline[linecolor=black, linewidth=0.08](12.58,4.02)(15.1,-0.24)(17.64,-4.56) \rput[bl](9.6,4.3){\mbox{\Huge $SU(N_c)$}} \rput[bl](15.64,-6.2){\mbox{\Huge $SU(N_F)_R$}} \psline[linecolor=black, linewidth=0.08](3.94,-4.54)(6.349498,-0.5963758)(9.04,4.04) \psframe[linecolor=black, linewidth=0.04, fillstyle=gradient, gradlines=2000, gradbegin=colour0, gradend=colour0, dimen=outer](8.8,-4.5)(0.0,-6.9) \rput[bl](2.2,-6.16){\mbox{\Huge $SU(N_F)_L$}} \psline[linecolor=black, linewidth=0.08, linestyle=dashed, dash=0.17638889cm 0.10583334cm](12.84,-5.68)(10.819074,-5.688932)(8.8,-5.68) \rput[bl](15.06,0.02){\mbox{\Huge $\tilde{Q}$}} \rput[bl](5.78,0.0){\mbox{\Huge ${Q}$}} \rput[bl](10.52,-5.48){\mbox{\Huge ${H}$}} \end{pspicture} }\end{center}\caption{\label{fig:Quiver-diagram-of}\emph{Quiver diagram of the fixed point
theory of \cite{Litim:2014uca}. Solid lines represent fermions, dashed
lines represent bosons.}}
\end{figure}

We begin by recapping the LS fixed point of \cite{Litim:2014uca}, whose
field content is shown in Table {\ref{LS - table},} where mid-alphabet
latin indices $i,j,k...$ are used to label flavour, while early-alphabet
latin indices $a,b,c$... label colour.  The particle content is represented as in a conventional quiver diagram, in Figure \ref{fig:Quiver-diagram-of}.
As usual, the circular nodes represents the $SU(N_{C})$ gauge factor, which is
crucial in establishing the LS fixed point. The
square nodes represent the flavour groups, $SU(N_{F})_{L}\otimes SU(N_{F})_{R}$,
which will become partially gauged in order to accommodate the electroweak
gauge factors of the SM. In \cite{Abel:2017ujy,Abel:2017rwl} the
LS model was augmented by coloured scalars in order to break the gauge
group down to the SM. However as mentioned in the Introduction,
there remain in such models light doublets which are
charged under the electroweak SM gauge groups. 

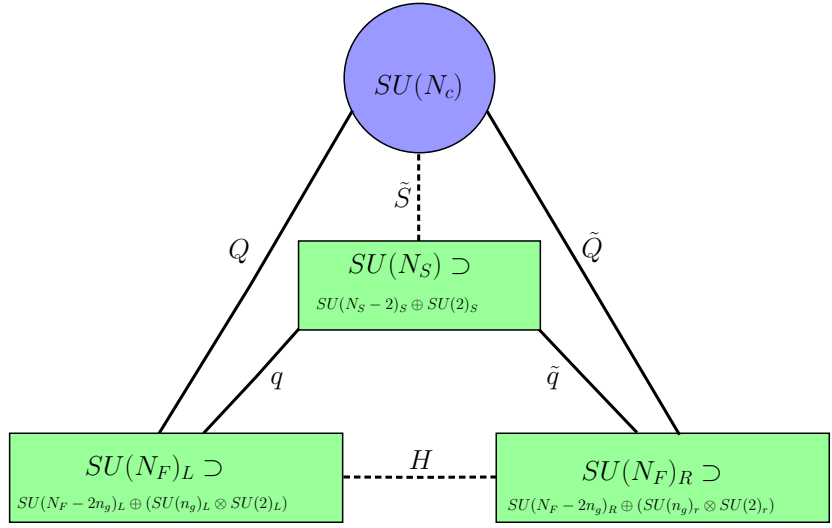
\begin{figure}
% \usepackage[usenames,dvipsnames]{pstricks} 
% \usepackage{epsfig} 
% \usepackage{pst-grad} % For gradients 
% \usepackage{pst-plot} % For axes 
% \usepackage[space]{grffile} % For spaces in paths 
% \usepackage{etoolbox} % For spaces in paths 
% \makeatletter % For spaces in paths 
% \patchcmd\Gread@eps{\@inputcheck#1 }{\@inputcheck"#1"\relax}{}{} 
% \makeatother % % User Packages: 
%  % 
\begin{center}
\psscalebox{0.5 0.5} % Change this value to rescale the drawing. 
{ \begin{pspicture}(0,-6.9)(21.6,6.9) \definecolor{colour4}{rgb}{0.2,0.8,0.0} \definecolor{colour6}{rgb}{0.6,1.0,0.6} \definecolor{colour5}{rgb}{0.4,0.4,1.0} \definecolor{colour3}{rgb}{0.6,0.6,1.0} \psframe[linecolor=black, linewidth=0.04, fillstyle=gradient, gradlines=2000, gradbegin=colour6, gradend=colour6, dimen=outer](21.6,-4.5)(12.8,-6.9) \pscircle[linecolor=black, linewidth=0.04, fillstyle=gradient, gradlines=2000, gradbegin=colour3, gradend=colour3, dimen=outer](10.8,4.9){2.0} \psline[linecolor=black, linewidth=0.08](12.58,4.02)(15.1,-0.24)(17.64,-4.56) \rput[bl](9.6,4.3){\mbox{\Huge $SU(N_c)$}} \rput[bl](13.12,-6.68){\mbox{\Large $SU(N_F-2n_g )_R\oplus (SU(n_g)_r\otimes SU(2)_r)$}} \rput[bl](15.06,-5.94){\mbox{\Huge $SU(N_F)_R\supset$}} \psline[linecolor=black, linewidth=0.08](3.94,-4.54)(6.349498,-0.5963758)(9.04,4.04) \psframe[linecolor=black, linewidth=0.04, fillstyle=gradient, gradlines=2000, gradbegin=colour6, gradend=colour6, dimen=outer](8.8,-4.5)(0.0,-6.9) \rput[bl](2.0,-5.8){\mbox{\Huge $SU(N_F)_L\supset$}} \rput[bl](0.22,-6.62){\mbox{\Large $SU(N_F-2n_g )_L\oplus (SU(n_g)_L\otimes SU(2)_L)$}} \psline[linecolor=black, linewidth=0.08](13.92,-1.74)(15.101031,-3.0006676)(16.52,-4.5) \psline[linecolor=black, linewidth=0.08](5.12,-4.52)(6.4907885,-3.0388641)(8.24,-1.06) \psframe[linecolor=black, linewidth=0.04, fillstyle=gradient, gradlines=2000, gradbegin=colour6, gradend=colour6, dimen=outer](14.0,0.6)(7.6,-1.8) \rput[bl](8.08,-1.34){\mbox{\Large $SU(N_S-2 )_S\oplus SU(2)_S$}} \rput[bl](8.92,-0.44){\mbox{\Huge $SU(N_S)\supset$}} \psline[linecolor=black, linewidth=0.08, linestyle=dashed, dash=0.17638889cm 0.10583334cm](10.78,2.88)(10.779074,1.7110679)(10.78,0.58) \psline[linecolor=black, linewidth=0.08, linestyle=dashed, dash=0.17638889cm 0.10583334cm](12.84,-5.68)(10.819074,-5.688932)(8.8,-5.68) \rput[bl](15.06,0.02){\mbox{\Huge $\tilde{Q}$}} \rput[bl](5.78,0.0){\mbox{\Huge ${Q}$}} \rput[bl](6.88,-3.36){\mbox{\Huge ${q}$}} \rput[bl](14.14,-3.34){\mbox{\Huge $\tilde{q}$}} \rput[bl](10.52,-5.48){\mbox{\Huge ${H}$}} \rput[bl](10.12,1.48){\mbox{\Huge $\tilde{S}$}} \end{pspicture} }\end{center}

\caption{\emph{The ``Tetrad'' quiver that gives the Standard Model in the
IR. Note that is not possible to illustrate the gauging of the electroweak
symmetries on such a diagram. On the right, the gauging is on the $SU(2)_{R}=[SU(2)_{r}\otimes SU(2)_{S}]_{{\diag}}$ factor, with 
the top $2n_{g}$ indices of $SU(N_{F})_{L,R}$ flavour transforming
as doublets under $SU(n_{g})\otimes SU(2)_{L,R}$. \label{fig:The-augmented-quiver}}}
\end{figure}

Let us now proceed directly to the phenomenologically viable augmented model
that we will propose in this paper. As we shall see the model leaves
no light states, other than those appearing directly in the SM, beyond
an enhanced Higgs sector (with the Higgs fields carrying the same generation indices as the matter fields). In this section we will lay out the spectrum
and pattern of VEVs that need to be achieved in order to realise the Standard
Model in the IR, and then in the following section we consider the
dynamics that achieves them. The augmented model is shown in Table
\ref{SM table} and its corresponding quiver diagram in Figure \ref{fig:The-augmented-quiver}.
It contains four elements, hence we refer to it as the Tetrad Model (TM$^{\mbox{\tiny TM}}$). 
As in \cite{Abel:2017rwl} it is an extension of the PS model to a larger unified group. Note that the PS gauge unification
to $SU(2)_{R}$ is adopted to take advantage of the $SU(2)$ large-flavour
fixed points, introduced in \cite{gracey,Pica:2010xq}. We will  use a
Weyl notation and display the left and right fermions explicitly.
We use the following nomenclature for the spectrum: Fermions
will be denoted with $Q$ and $q$'s, while scalars will be denoted
with $\tilde{S}$ and $H$'s\footnote{The $\tilde{S}$ scalars were referred to as $\tilde{Q}$ in \cite{Abel:2017rwl}
, but in the present context this would cause confusion.}. The flavour indices $i=1...N_{F}$ have three generations of components
gauged under electroweak $SU(2)_{L}$ and $SU(2)_{R}$. However we
have to gauge the right-handed component of the electroweak
gauge group in the correct way to yield the SM spectrum. Indeed the
``squarks'' $\tilde{S}$ have their own $SU(N_{S})$ flavour symmetry,
and the first two flavours also have to be charged under $SU(2)_{R}$
in order to give the correct PS breaking. The simplest solution is
then to identify \emph{$SU(2)_{R}=[SU(2)_{r}\otimes SU(2)_{S}]_{\diag}$}.
This leads to hypercharge $Y=\left(2T_{R}^{(3)}+B-L\right)$ and charge
$Q_{\mbox{\tiny \rm e.m.}}=\frac{1}{2}\left(2T_{R}^{(3)}+2T_{L}^{(3)}+B-L\right)$,
where $T_{L/R}^{(3)}=\text{diag}(\frac{1}{2},-\frac{1}{2})$ and $B-L$
is the $\text{diag}(\frac{1}{3},\,\frac{1}{3},\,\frac{1}{3},-1,\,0,\,0...,\,0)$
generator of $SU(N_{C})$. As we shall see, for the LS gauge-Yukawa fixed point to be weakly coupled we require 
$N_F\approx \frac{21}{4}N_C$. 

\begin{table}
\centering{}{\tiny{}}%
\caption{\emph{Fields in the arbitrarily weakly coupled asymptotic safe fixed
point of \cite{Litim:2014uca}} \label{LS - table}.}
\begin{tabular}{|c||c||c|c|c|}
\hline 
{\tiny{}$ $} & {\scriptsize{}$SU(N_{C})$} & {\scriptsize{}$SU(N_{F})_{L}$} & {\scriptsize{}$SU(N_{F})_{R}$} & spin\tabularnewline
\hline 
\hline 
$Q_{ai}$ & $\fund$ & $\fund$ & $1$ & $1/2$\tabularnewline
\hline 
$\tilde{Q}^{ia}$ & $\afund$ & $1$ & $\afund$ & $1/2$\tabularnewline
\hline 
$H_{j}^{i}$ & $1$ & $\afund$ & $\fund$ & 0\tabularnewline
\hline 
\end{tabular}
\end{table}

\begin{table}
\centering{}{\tiny{}}%
\caption{ \emph{Fields in the asymptotically safe ``Tetrad'' Model, where} $N_{S}=N_{C}-2$ and $N_F\approx \frac{21}{4}N_C$\emph{.
The top $2n_{g}=6$ components of flavour $SU(N_{F})$ correspond
to $SU(2)$ multiplets, where $n_{g}$ is the generation number}.
\emph{The gauging for the usual Pati-Salam
$SU(2)_{R}$ group is identified as $SU(2)_{R}=[SU(2)_{r}\otimes SU(2)_{S}]_{\mbox{\rm\tiny diag}}$.}}
\begin{tabular}{|c||c||c|c|c||c|}
\hline 
{\tiny{}$ $} & {\scriptsize{}$SU(N_{C})$} & {\scriptsize{}}%
\begin{tabular}{c}
{\scriptsize{}$SU(N_{F})_{L}\supset$}\tabularnewline
{\scriptsize{}$SU(2)_{L}\otimes SU(n_{g})_{L}$}\tabularnewline
\end{tabular} & {\scriptsize{}}%
\begin{tabular}{c}
{\scriptsize{}$SU(N_{F})_{R}\supset$}\tabularnewline
{\scriptsize{}$SU(2)_{r}\otimes SU(n_{g})_{r}$}\tabularnewline
\end{tabular} & {\scriptsize{}}%
\begin{tabular}{c}
{\scriptsize{}$SU(N_{S})=$}\tabularnewline
{\scriptsize{}$SU(N_{C}-4)_{S}\oplus SU(2)_{S}$}\tabularnewline
\end{tabular} & spin\tabularnewline
\hline 
\hline 
$Q_{ai}$ & $\fund$ & $\fund\supset(\fund,\fund)$ & $1$ & 1 & $1/2$\tabularnewline
\hline 
$\tilde{Q}^{ia}$ & $\afund$ & $1$ & $\afund\supset(\afund,\afund)$ & 1 & $1/2$\tabularnewline
\hline 
$H_{j}^{i}$ & $1$ & $\afund\supset(\afund,\afund)$ & $\fund\supset(\fund,\fund)$ & $1$ & 0\tabularnewline
\hline 
$\tilde{S}_{a,\ell=1..N_{S}}$ & $\afund$ & $1$ & $1$ & $\afund=\afund_{N_{C}-4}\oplus\afund_{2}$ & 0\tabularnewline
\hline 
\hline 
$\tilde{q}_{\ell}^{i}$ & 1 & $\afund\supset(\afund,\afund)$ & $1$ & $\fund=\fund_{N_{C}-4}\oplus\fund_{2}$ & $1/2$\tabularnewline
\hline 
$q_{j}^{\ell}$ & 1 & $1$ & $\fund\supset(\fund,\fund)$ & $\afund=\afund_{N_{C}-4}\oplus\afund_{2}$ & $1/2$ 
\tabularnewline
\hline
\end{tabular}
\end{table}
The necessity of the additional fermionic fields $q,\tilde{q}$ can be deduced from the requirement that they are 
able to remove the unwanted
light fermionic degrees of freedom while maintaining the chiral symmetry. The allowed couplings one can consider for the generation of the UV-fixed point are
\begin{align}
\mathcal{L}_{\mbox{\tiny UVFP}} & ~\supset~\mathcal{L}_{\mbox{\tiny KE}}+\frac{y}{\sqrt{2}}\text{Tr}\left[\left(QH\right)\cdot\tilde{Q}\right]+\frac{\tilde{y}}{\sqrt{2}}\text{Tr}\left[qH^{\dagger}\tilde{q}\right]-\frac{\tilde{Y}}{\sqrt{2}}\text{Tr}[\left(\tilde{S}\cdot Q\right)\tilde{q}]-\frac{Y}{\sqrt{2}}\text{Tr}[\left(\tilde{Q}\cdot\tilde{S}^{\dagger}\right)q]\nonumber \\
 & \,\,\,\,\,\,\qquad\qquad\,\,\qquad\qquad-u_{1}\text{Tr}\left[H^{\dagger}H\right]^{2}-u_{2}\text{Tr}\left[H^{\dagger}H\,H^{\dagger}H\right]-v_{1}\text{Tr}\left[H^{\dagger}H\right]\text{Tr}\left[\tilde{S}^{\dagger}\cdot\tilde{S}\right]\nonumber \\
 & \,\,\,\,\,\,\,\,\,\,\,\,\,\,\,\,\,\,\qquad\qquad\qquad\qquad\qquad\qquad\qquad\qquad-w_{1}\text{Tr}\left[\tilde{S}^{\dagger}\cdot\tilde{S}\right]^{2}-w_{2}\text{Tr}\left[\tilde{S}^{\dagger}\cdot\tilde{S}\,\tilde{S}^{\dagger}\cdot\tilde{S}\right]\,\, ,\label{eq:Luvfp}
\end{align}
 where the trace is over the flavour indices and the dot refers to colour
contraction. As we shall see the $Y$ and $\tilde{Y}$ Yukawa couplings are responsible for giving
masses to the unwanted degrees of freedom in the IR once $\tilde{S}$ gets a VEV. They are written
above somewhat schematically as clearly they cannot couple all the
flavour components in the same way due to the $SU(2)_{R}$ gauge invariance.
They will be treated explicitly below.

As in \cite{Abel:2017rwl} we will not consider the flavour breaking coupling (schematically)
\begin{equation}
\mathcal{L}_{\mbox{\tiny $\cancel{SU(N_{F})}$ }}~=~-v_{2}\text{Tr}\left[H^{\dagger}H\,\tilde{S}^{\dagger}\cdot\tilde{S}\right]\,\,\,.
\end{equation}
This coupling can be fixed to be precisely zero, where it will remain along the flow. 
(It can of course be forbidden on grounds of preservation of flavour
symmetry which we will associate with the classically relevant operators
only.) As we shall see the flavour conserving portal coupling $v_1$ can generate a mass-squared for the electroweak Higgses, and we keep it in the analysis.

We can in addition include the aforementioned dimensionful
``soft-terms''. Unlike the classically dimensionless couplings these will be allowed to explicitly violate the flavour symmetry. They
can be written most generally in the form 
\begin{equation}
\mathcal{L}_{\mbox{\tiny{Soft}}}\,\,=\,\,-m_{h_{0}}^{2}\text{Tr}\left[H^{\dagger}H\right]-\sum_{a=1}^{N_{F}^{2}-1}\Delta_{a}^{2}\text{Tr}\left[HT^{a}\right]\text{Tr}\left[H^{\dagger}T^{a}\right]\,\,,\label{eq:lsoft}
\end{equation}
where $T^{a}$ are the generators of the $SU(N_{F})_{{\diag}}$ flavour
group. Being classically relevant, the soft terms cannot disrupt the
UV fixed point, but can serve to generate symmetry breaking themselves, and also remove any Goldstone modes associated
with the spontaneously broken global flavour symmetries.

\subsection{\it Structure in the IR -- emergence of the SM from the TM}

Next let us confirm that the SM emerges in the IR from the Tetrad Model. 
It is useful for this purpose to explicitly write the particle content
in terms of SM quantum numbers in order to discuss the couplings,
and determine the required values for $N_{F},N_{S}$: the 
explicit representations are (c.f. the usual PS model in for example \cite{King:1997ia}) 
\begin{align}
\,\,\, & \:\,\,\,\,\,\,\,\,\,\,\,\,\,\,\,\,\,\,\,\,\,\,\,\,\,\,\,\,\,\,\,\,\,\,\,\,\,\,\,\,\,\,\,\,\,\,\,N_{C}\nonumber \\
Q\,\, & \,\,\,=\,\,\left.\left(\overbrace{\begin{array}{ccc}
q_{1} & \ell_{1} & \cdots\left(\begin{array}{c}
\Psi_{\frac{1}{2}}\\
\Psi_{-\frac{1}{2}}
\end{array}\right)\cdots\\
q_{2} & \ell_{2} & \cdots\left(\begin{array}{c}
\Psi_{\frac{1}{2}}\\
\Psi_{-\frac{1}{2}}
\end{array}\right)\cdots\\
q_{3} & \ell_{3} & \cdots\left(\begin{array}{c}
\Psi_{\frac{1}{2}}\\
\Psi_{-\frac{1}{2}}
\end{array}\right)\cdots\\
\vdots & \vdots & \ddots
\end{array}}\right)\right\} N_{F}\,\,\,\,;\,\,\,\,\,\,\tilde{Q}=\left(\begin{array}{ccc}
\left(\begin{array}{c}
u^{c}\\
d^{c}
\end{array}\right) & \left(\begin{array}{c}
\nu_{e}^{c}\\
e^{c}
\end{array}\right) & \cdots\left(\begin{array}{c}
\tilde{\Psi}_{-\frac{1}{2}}\\
\tilde{\Psi}_{\frac{1}{2}}
\end{array}\right)\cdots\\
\left(\begin{array}{c}
s^{c}\\
c^{c}
\end{array}\right) & \left(\begin{array}{c}
\nu_{\mu}^{c}\\
\mu^{c}
\end{array}\right) & \cdots\left(\begin{array}{c}
\tilde{\Psi}_{-\frac{1}{2}}\\
\tilde{\Psi}_{\frac{1}{2}}
\end{array}\right)\cdots\\
\left(\begin{array}{c}
b^{c}\\
t^{c}
\end{array}\right) & \left(\begin{array}{c}
\nu_{\tau}^{c}\\
\tau^{c}
\end{array}\right) & \cdots\left(\begin{array}{c}
\tilde{\Psi}_{-\frac{1}{2}}\\
\tilde{\Psi}_{\frac{1}{2}}
\end{array}\right)\cdots\\
\vdots & \vdots & \ddots
\end{array}\right)\,\,\,,\label{eq:matter}
\end{align}
\begin{align}
\,\,\, & \:\,\,\,\,\,\,\,\,\,\,\,\,\,\,\,\,\,\,\,\,\,\,\,\,\,\,\,\,\,\,\,\,\,\,\,\,\,\,\,\,N_{S}=N_{C}-2\nonumber \\
q\,\, & \,\,\,=\,\,\left.\left(\overbrace{\begin{array}{ccc}
\left(\begin{array}{cc}
\psi_{0} & \psi_{1}\\
\psi_{-1} & \psi_{0}
\end{array}\right)\left(\begin{array}{c}
\psi_{\frac{1}{2}}\\
\psi_{-\frac{1}{2}}
\end{array}\right) & \dots & \left(\begin{array}{c}
\psi_{\frac{1}{2}}\\
\psi_{-\frac{1}{2}}
\end{array}\right)\\
\left(\begin{array}{cc}
\psi_{0} & \psi_{1}\\
\psi_{-1} & \psi_{0}
\end{array}\right)\left(\begin{array}{c}
\psi_{\frac{1}{2}}\\
\psi_{-\frac{1}{2}}
\end{array}\right) & \dots & \left(\begin{array}{c}
\psi_{\frac{1}{2}}\\
\psi_{-\frac{1}{2}}
\end{array}\right)\\
\left(\begin{array}{cc}
\psi_{0} & \psi_{1}\\
\psi_{-1} & \psi_{0}
\end{array}\right)\left(\begin{array}{c}
\psi_{\frac{1}{2}}\\
\psi_{-\frac{1}{2}}
\end{array}\right) & \dots & \left(\begin{array}{c}
\psi_{\frac{1}{2}}\\
\psi_{-\frac{1}{2}}
\end{array}\right)\\
\vdots & \vdots & \vdots
\end{array}}\right)\right\} N_{F}\,\,\,\,;\,\,\,\,\,\,\,\tilde{q}\,\,=\,\,\left(\begin{array}{ccc}
\left(\begin{array}{c}
\tilde{\psi}_{-\frac{1}{2}}\\
\tilde{\psi}_{\frac{1}{2}}
\end{array}\right) & \dots & \left(\begin{array}{c}
\tilde{\psi}_{-\frac{1}{2}}\\
\tilde{\psi}_{\frac{1}{2}}
\end{array}\right)\\
\left(\begin{array}{c}
\tilde{\psi}_{-\frac{1}{2}}\\
\tilde{\psi}_{\frac{1}{2}}
\end{array}\right) & \dots & \left(\begin{array}{c}
\tilde{\psi}_{-\frac{1}{2}}\\
\tilde{\psi}_{\frac{1}{2}}
\end{array}\right)\\
\left(\begin{array}{c}
\tilde{\psi}_{-\frac{1}{2}}\\
\tilde{\psi}_{\frac{1}{2}}
\end{array}\right) & \dots & \left(\begin{array}{c}
\tilde{\psi}_{-\frac{1}{2}}\\
\tilde{\psi}_{\frac{1}{2}}
\end{array}\right)\\
\vdots & \vdots & \vdots
\end{array}\right)\,\,\,,
\end{align}
\begin{align}
\,\,\, & \:\,\,\,\,\,\,\,\,\,\,\,\,\,\,\,\,\,\,\,\,\,\,\,\,\,\,\,\,\,\,\,\,\,\,\,\,\,\,\,\,\,\,\,\,\,\,\,\,\,\,\,\,\,\,\,\,\,\,\,\,\,\,\,\,\,\,\,\,\,\,\,\,\,\,\,\,\,\,N_{C}\nonumber \\
\tilde{S}\,\, & \,\,\,=\,\,\left(\begin{array}{c}
S_{PS}\\
\Phi_{0}
\end{array}\right)\,\,=\,\,\left.\left(\overbrace{\begin{array}{ccccc}
\left(\begin{array}{c}
\tilde{d}^{c}\\
\tilde{u}^{c}
\end{array}\right) & \left(\begin{array}{c}
\tilde{e}^{c}\\
\tilde{\nu}^{c}
\end{array}\right) & \left(\begin{array}{c}
\tilde{\phi}_{-\frac{1}{2}}\\
\tilde{\phi}_{\frac{1}{2}}
\end{array}\right) & \cdots & \left(\begin{array}{c}
\tilde{\phi}_{-\frac{1}{2}}\\
\tilde{\phi}_{\frac{1}{2}}
\end{array}\right)\\
\tilde{T}_{-\frac{1}{6}} & \tilde{\phi}_{\frac{1}{2}} & \tilde{\phi}_{0} & \ldots & \tilde{\phi}_{0}\\
\vdots & \vdots & \vdots &  & \vdots\\
\tilde{T}_{-\frac{1}{6}} & \tilde{\phi}_{\frac{1}{2}} & \tilde{\phi}_{0} & \ldots & \tilde{\phi}_{0}
\end{array}}\right)\right\} \,N_{S}=N_{C}-2\,\,\,\,,\label{eq:S-field}
\end{align}
\begin{equation}
H=\left(\begin{array}{cccc}
\left(\begin{array}{cc}
h_{u}^{0} & h_{d}^{-}\\
h_{u}^{+} & h_{d}^{0}
\end{array}\right)_{11} & \left(\begin{array}{cc}
h_{u}^{0} & h_{d}^{-}\\
h_{u}^{+} & h_{d}^{0}
\end{array}\right)_{12} & \left(\begin{array}{cc}
h_{u}^{0} & h_{d}^{-}\\
h_{u}^{+} & h_{d}^{0}
\end{array}\right)_{13} & \cdots\\
\left(\begin{array}{cc}
h_{u}^{0} & h_{d}^{-}\\
h_{u}^{+} & h_{d}^{0}
\end{array}\right)_{21} & \left(\begin{array}{cc}
h_{u}^{0} & h_{d}^{-}\\
h_{u}^{+} & h_{d}^{0}
\end{array}\right)_{22} & \left(\begin{array}{cc}
h_{u}^{0} & h_{d}^{-}\\
h_{u}^{+} & h_{d}^{0}
\end{array}\right)_{23} & \cdots\\
\left(\begin{array}{cc}
h_{u}^{0} & h_{d}^{-}\\
h_{u}^{+} & h_{d}^{0}
\end{array}\right)_{31} & \left(\begin{array}{cc}
h_{u}^{0} & h_{d}^{-}\\
h_{u}^{+} & h_{d}^{0}
\end{array}\right)_{32} & \left(\begin{array}{cc}
h_{u}^{0} & h_{d}^{-}\\
h_{u}^{+} & h_{d}^{0}
\end{array}\right)_{33} & \cdots\\
\vdots & \vdots & \vdots & H_{0}
\end{array}\right)\,\,,
\end{equation}
where $H_{0}$ is an $\left(N_{F}-6\right)\times\left(N_{F}-6\right)$
scalar which is uncharged under the SM gauge groups, and the sufficies
denote $Q_{\mbox{\tiny \rm e.m.}}$. The assignment of the remaining fields is obvious.

First note that the top $2n_{g}$ (where $n_{g}=3$ is the number of generations, but 
it is often useful to leave it generic)
entries of flavour are charged under the $SU(2)$ gauge groups. Therefore,
given the couplings and matter content, there are $n_{g}$ generations
of SM Higgs doublets in the top $2n_{g}\times2n_{g}$ components of
$H$. Assuming that $n_{g}=3$, this corresponds
to 18 separate Higgs $SU(2)_{L}$ doublets. Clearly one ultimately
requires these to be lifted in a hierarchical way so that there is
one dominant lighter Higgs which gets a VEV, which will be a mixture
of the 18 original ones. 
In contrast with \cite{Abel:2017rwl} we will assume
that the scalars $\tilde{S}$ are gauged only under colour except
for the first two flavours which are charged under the gauged $SU(2)_{R}$. (The latter choice is flexible.)

We repeat that we are assuming flavour degeneracy in all the couplings
of (\ref{eq:Luvfp}). One could instead for example take the $y$
Yukawa couplings to break $SU(N_{F})_{L}\times SU(N_{F})_{R}$ symmetry,
but this would require a  re-analysis of the UV fixed point
behaviour of the theory so we instead adopt the philosophy of \cite{Abel:2017rwl}. 
As there
is a pair of Higgs multiplets for each generation, this is indeed
an attractive possibility for introducing SM-flavour structure. Moreover
as shown in \cite{Abel:2017rwl} and expanded upon below, the flavour universal
part of such operators flows to relatively smaller absolute values, ``exposing'' flavour hierarchies during the flow, so that they become
dominant in the IR. 

There are two elements to the gauge symmetry breaking. First there
are VEVs for  $\tilde{S}$. We must choose $N_{S}=N_{C}-2$,
so that they can be rearranged by suitable colour and $SU(N_{S})$
flavour rotations into the form 

\begin{align}
\,\,\, & \:\,\,\,\,\,\,\,\,\,\,\,\,\,\,\,\,\,\,\,\,\,\,\,\,\,\,\,\,\,\,\,\,\,\,\,\,\,\,\,\,\,\,\,\,\,\,N_{C}\nonumber \\
\langle\tilde{S}\rangle\,\, & =\,\,\tilde{V}\left.\left(\overbrace{\begin{array}{cccccc}
0 & 0 & 0 & 0 & \cdots & 0\\
\vdots & \vdots & \vdots & 1 &  & \vdots\\
\vdots & \vdots & \vdots &  & \ddots & \vdots\\
0 & 0 & 0 & 0 & \cdots & 1
\end{array}}\right)\right\} N_{S}=N_{C}-2\,\,\,,
\end{align}
with the VEV $\tilde{\phi}_{0}$ in (\ref{eq:S-field}) being of the
form $\langle\tilde{\phi}_{0}\rangle=\tilde{V}\mathbb{I}_{N_{C-4}}$,
where $\tilde{V}$ is a constant. The $SU(2)_{R}$ orientation simply
determines the direction corresponding to the massless right-handed
``sneutrino'', so one may always choose a basis in which the $\tilde{\nu}^{c}$
and $N_{C}-4$ of the $\tilde{\phi}_{0}$'s on the diagonal get a
VEV. (Obviously the case $N_{C}=4$ is the standard, non-asymptotically
free, Pati-Salam model.)

 At this stage the gauge symmetry is broken
to the Standard Model as 
\begin{equation}
SU(N_{C})\times SU(2)_{L}\times SU(2)_{R}~\longrightarrow~ SU(3)_{c}\times SU(2)_{L}\times U(1)_{Y}\,\,.
\end{equation}
Given that the gauge symmetry can be broken as required, one can 
focus on the excess states that need to be made massive
in order to end up with the Standard Model in the IR. In particular
there are of course (by design) very many $SU(2)_{L}$ and $SU(2)_{R}$
doublets that should be removed at low scales. The second component
of symmetry breaking that accomplishes this is that the block $H_{0}$
of the Higgs multiplets also acquire VEV along the diagonal, 
\begin{equation}
\langle H_{0}\rangle\,\,=\,\,V_{0}\mathbb{I}_{N_{F}-6}\,.
\end{equation}
Thanks to the $y$ coupling, this gives the $N_{F}-6$ generations
of complete non-doublet $SU(N_{C})$ multiplets masses $\frac{yV_{0}}{\sqrt{2}}$,
leaving untouched $n_{g}(N_{C}-4)$ of the $SU(2)_{L}$ doublets in
the $Q_{L}$, and $SU(2)_{R}$ doublets in the $Q_{R}$. Indeed in
these remaining $n_{g}$ generations of $SU(N_{C})$-coloured multiplets,
only the first $SU(4)$ components are to be identified as matter
fields, as in (\ref{eq:matter}). The remaining states get masses
$\frac{\tilde{Y}\tilde{V}}{\sqrt{2}}$ and $\frac{Y\tilde{V}}{\sqrt{2}}$
from the $\tilde{Y}$ and $Y$ couplings respectively to which we
now return, writing them with explicit indices: 
\begin{align}
\mathcal{L}_{\mbox{\tiny \rm UVFP}} & \supset-\frac{\tilde{Y}}{\sqrt{2}}Q\tilde{S}\tilde{q}-\frac{Y}{\sqrt{2}}\tilde{S}^{\dagger}\tilde{Q}q\,\,\,,\nonumber \\
 & \supset-\frac{\tilde{Y}}{\sqrt{2}}\left(Q_{k}^{a\alpha}\tilde{S}_{a}^{j}\tilde{q}_{\alpha j}^{k}\right)-\frac{Y}{\sqrt{2}}\left(\tilde{Q}_{k}^{a\alpha}S_{a}^{*j}q_{\alpha j}^{k}\right)\,\,\,,\nonumber \\
 & \equiv-\frac{\tilde{Y}}{\sqrt{2}}\left(\begin{array}{cc}
\Psi_{\frac{1}{2}} & \Psi_{-\frac{1}{2}}\end{array}\right)_{k}^{\hat{a}}\tilde{\phi}_{0,\hat{a}}^{j}\left(\begin{array}{c}
\tilde{\psi}_{-\frac{1}{2}}\\
\tilde{\psi}_{\frac{1}{2}}
\end{array}\right)_{j}^{k}-\frac{Y}{\sqrt{2}}\left(\begin{array}{cc}
\tilde{\Psi}_{-\frac{1}{2}} & \tilde{\Psi}_{\frac{1}{2}}\end{array}\right)_{k}^{\hat{a}}\tilde{\phi}_{0,\hat{a}}^{*j}\left(\begin{array}{c}
\psi_{\frac{1}{2}}\\
\psi_{-\frac{1}{2}}
\end{array}\right)_{j}^{k}\,\,\,,\nonumber \\
 & =-\frac{\tilde{Y}\tilde{V}}{\sqrt{2}}\left(\begin{array}{cc}
\Psi_{\frac{1}{2}} & \Psi_{-\frac{1}{2}}\end{array}\right)_{k}^{j}\left(\begin{array}{c}
\tilde{\psi}_{-\frac{1}{2}}\\
\tilde{\psi}_{\frac{1}{2}}
\end{array}\right)_{j}^{k}-\frac{Y\tilde{V}}{\sqrt{2}}\left(\begin{array}{cc}
\tilde{\Psi}_{-\frac{1}{2}} & \tilde{\Psi}_{\frac{1}{2}}\end{array}\right)_{k}^{j}\left(\begin{array}{c}
\psi_{\frac{1}{2}}\\
\psi_{-\frac{1}{2}}
\end{array}\right)_{j}^{k}\,\,\,,
\end{align}
where $a=1\ldots N_{C}$ are colour indices, $\hat{a}=5\ldots N_{C}$
are the $N_{C}-4$ colour indices beyond the PS degrees of freedom,
$j=1\ldots N_{C}-4$ are the $N_{C}-4$ flavour indices of $\tilde{S}$
that are not charged under $SU(2)_{R}$, the indices $\alpha=1,2$
are the $SU(2)_{L/R}$ indices, and $k=1\ldots n_{g}$ are generation
indices. Note that as promised
chiral symmetry dictates the choice $N_{S}=N_{C}-4$, because 
$N_{S}$ flavour is locked to  $SU(N_{C})$ colour by the VEV of
$\tilde{S}$. 

It is easy to check that with this choice of colours and flavours, and these VEVs, 
the remaining content in the IR is that of the SM with Higgses carrying $SU(n_g)$ generation 
indices for the left and right handed fields.

\section{Flow from the UV fixed points and symmetry breaking}
\subsection{\it The Tetrad Model contains the Coleman Weinberg mechanism}

Next let us turn to the dynamics, first illustrating the appearance of traditional radiative symmetry breaking. 
As there are many couplings involved, it is useful to break down the evolution under RG flow into self-contained 
units. Indeed the crucial aspect of the flow from the UV fixed point is 
that it is actually controlled by {\it two} fixed points of the gauge and
Yukawa couplings,
which form a closed system by themselves. 

It will be convenient to define rescaled couplings as follows:
\begin{align}
\alpha_{g} & \,\,=\,\,\frac{N_{C}g^{2}}{(4\pi)^{2}}\,;\,\,\alpha_{y}=\frac{N_{C}y^{2}}{(4\pi)^{2}};\,\,\alpha_{\tilde{y}}=\frac{N_{C}\tilde{y}^{2}}{(4\pi)^{2}};\,\,\alpha_{Y}=\frac{N_{C}Y^{2}}{(4\pi)^{2}};\,\,\alpha_{\tilde{Y}}=\frac{N_{C}\tilde{Y}^{2}}{(4\pi)^{2}}\,;\nonumber \\
\alpha_{u_{1}} & \,\,=\,\,\frac{N_{F}^{2}u_{1}}{(4\pi)^{2}}\,;\,\,\alpha_{u_{2}}=\frac{N_{F}u_{2}}{(4\pi)^{2}}\,;\,\,\alpha_{v_{1}}=\frac{N_{C}^{2}v_{1}}{(4\pi)^{2}};\,\,\alpha_{w_{1}}=\frac{N_{C}^{2}w_{1}}{(4\pi)^{2}}\,;\,\,\alpha_{w_{2}}=\frac{N_{C}w_{2}}{(4\pi)^{2}}\,.
\end{align}
To determine their fixed points, we require
their RG equations to order $\alpha^{3}\equiv\epsilon\alpha^{2}$
in $\beta_{g}$ and order $\alpha^{2}\equiv\epsilon\alpha$ in $\beta_{y,Y,\tilde{Y}}$:
defining $\epsilon=-11/2+x_{F}+x_{q}/4=x_{F}-21/4$ and $\Upsilon=\sqrt{\alpha_{y}\alpha_{\tilde{y}}\alpha_{Y}\alpha_{\tilde{Y}}}$,
and taking $n_{g}=3$, the beta functions are found to be  
\begin{align}
\beta_{g} & ~=~\alpha_{g}^{2}\left(\frac{4}{3}\epsilon+(\frac{26}{3}x_{F}-20)\alpha_{g}-x_{F}^{2}\alpha_{y}-x_{F}\alpha_{Y}-x_{F}\alpha_{\tilde{Y}}\right)\,\,,\nonumber \\
\beta_{y} & ~=~4\Upsilon+\alpha_{y}\left((1+x_{F})\alpha_{y}+\alpha_{\tilde{y}}+\alpha_{\tilde{Y}}+\alpha_{Y}-6\alpha_{g}\right)\,\,,\nonumber \\
\beta_{\tilde{y}} & ~=~4\Upsilon+\alpha_{\tilde{y}}\left((1+x_{F})\alpha_{\tilde{y}}+\alpha_{y}+\alpha_{\tilde{Y}}+\alpha_{Y}\right)\,\,,\nonumber \\
\beta_{Y} & ~=~2x_{F}\Upsilon+\alpha_{Y}\left(2(1+x_{F})\alpha_{Y}+x_{F}(\frac{1}{2}\alpha_{y}+\frac{1}{2}\alpha_{\tilde{y}}+2\alpha_{\tilde{Y}})-3\alpha_{g}\right)\,\,,\nonumber \\
\beta_{\tilde{Y}} & =2x_{F}\Upsilon+\alpha_{\tilde{Y}}\left(2(1+x_{F})\alpha_{\tilde{Y}}+x_{F}(\frac{1}{2}\alpha_{y}+\frac{1}{2}\alpha_{\tilde{y}}+2\alpha_{Y})-3\alpha_{g}\right)\,\,.
\end{align}
Since it is positive, the equation for $\beta_{\tilde{y}}=0$ can only
be consistently met with $\alpha_{\tilde{y}}=\Upsilon=0$. Moreover
if any of the other couplings are non-zero, it flows to zero in the
IR, so along the RG trajectory from any eligible fixed point it must
remain zero. There are also by inspection no positive solutions with
$\alpha_{y}=0$. In addition the last two equations allow a fixed
point if $\alpha_{Y}=\alpha_{\tilde{Y}}$ or one or both couplings
vanish. Hence one finds the possible flows shown in Table \ref{tab:The-collection-of} (where $\alpha_{g}^{*}$
is the fixed point value of the gauge coupling, taking $x_{F}\rightarrow21/4$).
\begin{table}
\centering{}%
\caption{\emph{The collection of UV fixed points for the gauge and Yukawa couplings:
schematically the flow is from $\text{A}\rightarrow\text{B}\rightarrow\text{C,D}\rightarrow\text{E}.$
Fixed points C,D,E are pseudo-fixed points in the sense that the quartic
scalar coupings do not have a fixed point there. The only true non-trivial
fixed point is the LS fixed point of B. \label{tab:The-collection-of}}}
\begin{tabular}{|c|c|c|c|c|c|}
\hline 
Label & $\alpha_{g}^{*}$ & $\alpha_{\tilde{y}}/\alpha_{g}$ & $\alpha_{y}/\alpha_{g}$ & $\alpha_{Y}/\alpha_{g}$ & $\alpha_{\tilde{Y}}/\alpha_{g}$\tabularnewline
\hline 
\hline 
A & $0$ & $0$ & $0$ & $0$ & $0$\tabularnewline
\hline 
B & $\frac{25}{18}\epsilon$ & $0$ & $\frac{6}{1+x_{F}}\rightarrow\,\frac{24}{25}$ & $0$ & $0$\tabularnewline
\hline 
C & $\frac{302}{225}\epsilon$ & $0$ & $\frac{6(3+4x_{F})}{4+7x_{F}+4x_{F}^{2}}\rightarrow\,\frac{144}{151}$ & $\frac{6}{4+7x_{F}+4x_{F}^{2}}\rightarrow\,\frac{6}{151}$ & $0$\tabularnewline
\hline 
D & $\frac{302}{225}\epsilon$ & $0$ & $\frac{6(3+4x_{F})}{4+7x_{F}+4x_{F}^{2}}\rightarrow\,\frac{144}{151}$ & $0$ & $\frac{6}{4+7x_{F}+4x_{F}^{2}}\rightarrow\,\frac{6}{151}$\tabularnewline
\hline 
E & $\frac{277}{207}\epsilon$ & 0 & $\frac{6(1+4x_{F})}{2+5x_{F}+4x_{F}^{2}}\rightarrow\,\frac{264}{277}$ & $\frac{3}{2+5x_{F}+4x_{F}^{2}}\rightarrow\,\frac{6}{277}$ & $\frac{3}{2+5x_{F}+4x_{F}^{2}}\rightarrow\,\frac{6}{277}$\tabularnewline
\hline 
\end{tabular}
\end{table}

\begin{figure}
\centering{}\includegraphics[scale=0.5]{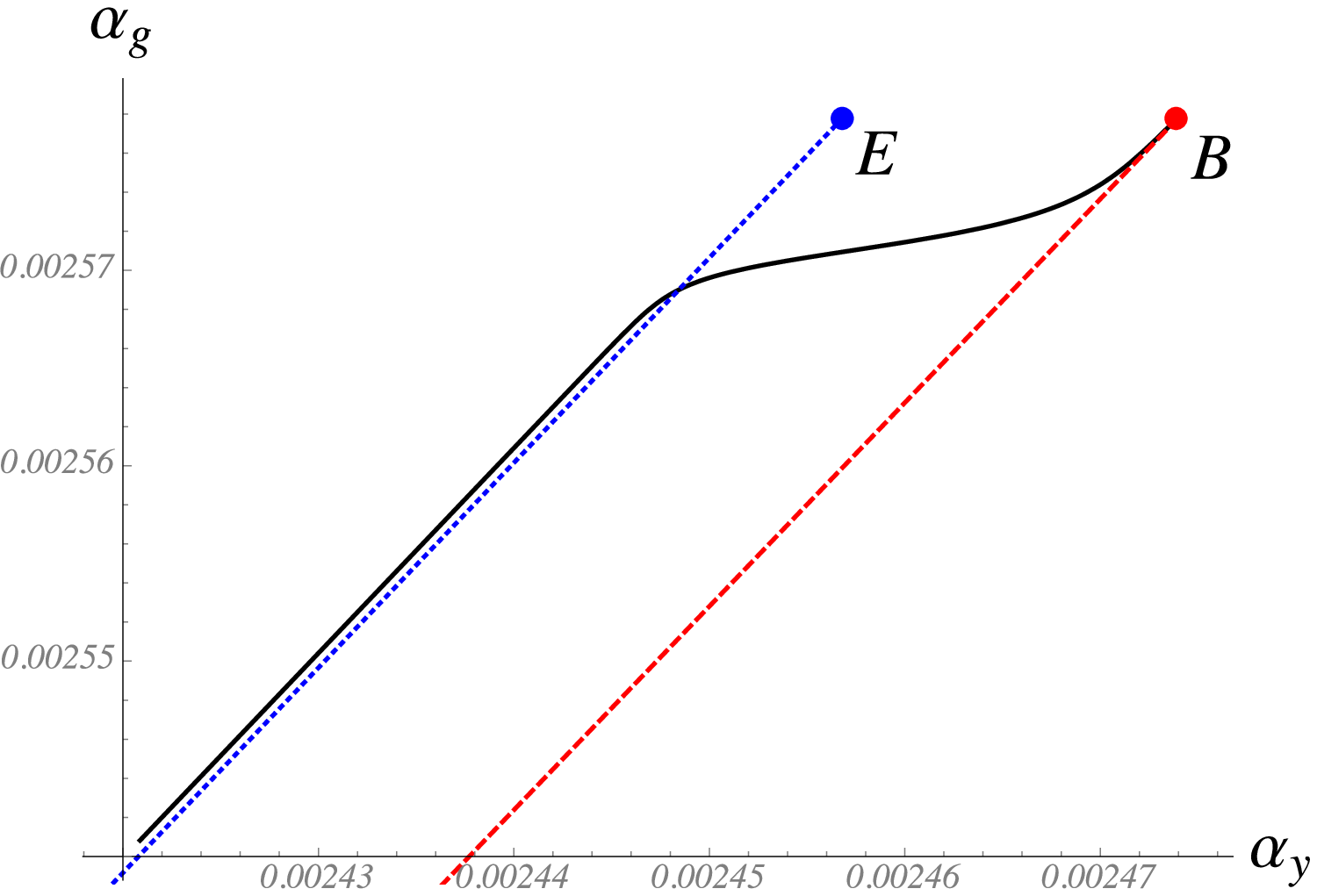}\includegraphics[scale=0.5]{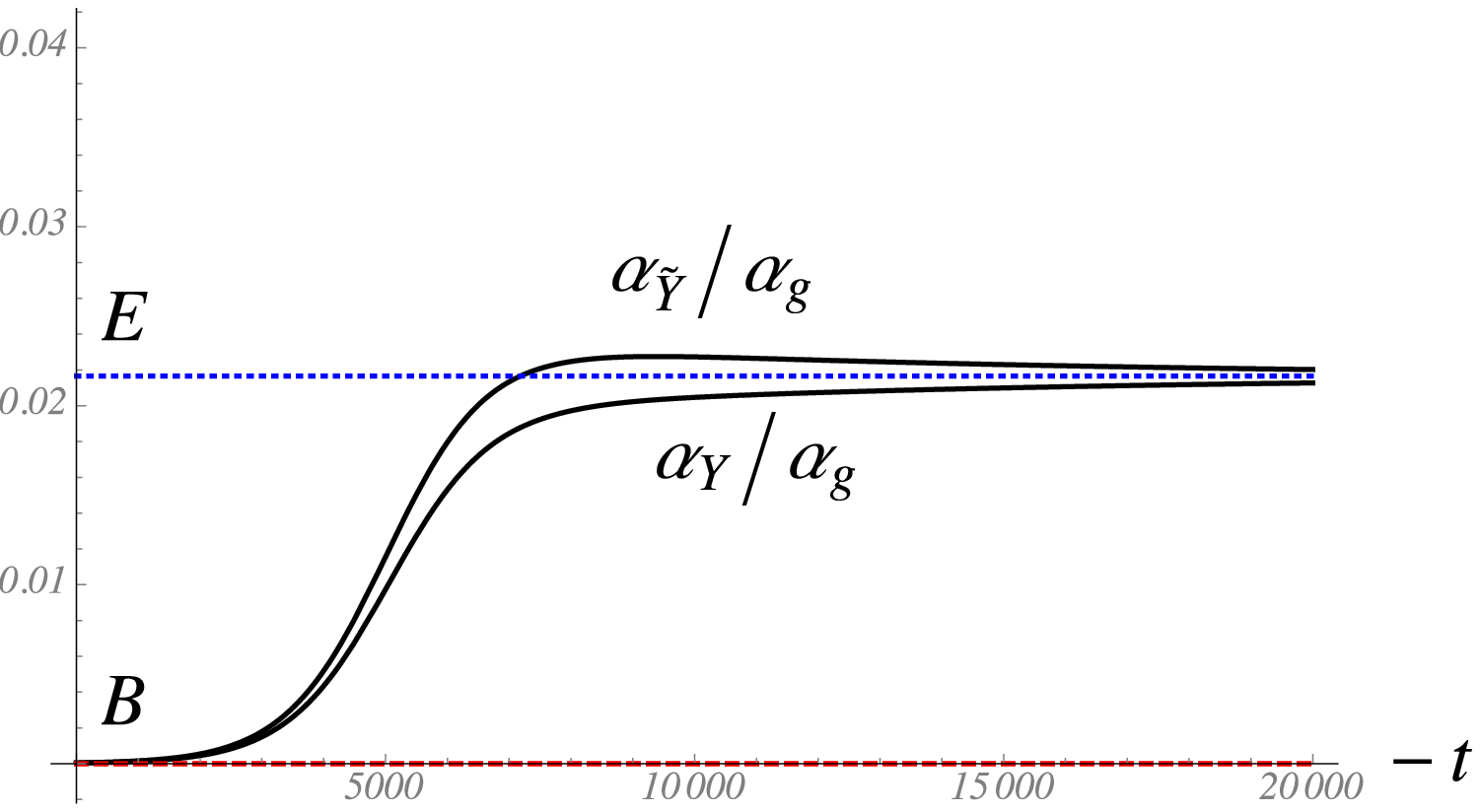}\caption{\emph{The 
 flow in gauge and Yukawa coupling-space from the true fixed
point B on to the trajectory emanating from the pseudo-fixed E, as
specified in Table \ref{tab:The-collection-of}. During the flow the
system crosses over from the (red-dashed) trajectory emanating from
B, onto the blue-dotted trajectory emanating from E, inducing non-zero
$\alpha_{Y},\alpha_{\tilde{Y}}$. This evolution (which ignores
the accompanying flow of the quartic couplings) is idealised:
the system radiatively develops a minimum before reaching trajectory
E. \label{fig:The-flow-in}}}
\end{figure}

Note that we do not require all the couplings to be non-zero in order
to have a non-trivial UV fixed point, but we definitely need to reproduce
the gauge-Yukawa behaviour of \cite{Litim:2014uca} that was observed in \cite{Abel:2017rwl},
while at the same time having negative beta functions for the couplings
that are required to be non-zero in the IR, for phenomenological reasons.
Therefore we can reject the Gaussian fixed point A. The second of
these options, fixed point B, was the LS fixed point that was utilized
in \cite{Abel:2017rwl}, and leads to 
\begin{equation}
\text{B}:\,\,\,\frac{\beta_{Y}}{\alpha_{Y}}=\frac{\beta_{\tilde{Y}}}{\alpha_{\tilde{Y}}}\approx-\frac{3}{1+x_{F}}\alpha_{g}\,\rightarrow-\frac{12}{25}\alpha_{g}\,<\,0\,\,\,,
\end{equation}
so that both the $Y$ and $\tilde{Y}$ couplings flow away from fixed
point B in the IR. Hence this fixed point is an interesting
Gaussian option for the $Y$ and $\tilde{Y}$ couplings. In order
to assess the other possible fixed points, note that
\begin{equation}
\frac{\beta_{Y}-\beta_{\tilde{Y}}}{\alpha_{Y}-\alpha_{\tilde{Y}}}\,\,=\,\,2(1+x_{F})\left(\alpha_{Y}+\alpha_{\tilde{Y}}\right)\,>0\,\,.
\end{equation}
Hence $\alpha_{Y}-\alpha_{\tilde{Y}}$ shrinks in the IR, so if the
flow begins in the UV at C or D, it will be attracted to fixed point
E. We conclude that from the perspective of the couplings $g,y,\tilde{y},Y,\tilde{Y}$,
any of B,C,D,E are suitable for an asympotically safe fixed point
but, as it flows to the IR, the system is attracted to the trajectory
emerging from fixed point E, driven by the Yukawa couplings $Y,\tilde{Y}$. 
A numerical evolution showing this cross-over for the Yukawa and gauge couplings
is shown in Figure \ref{fig:The-flow-in}.

Next we turn to  the scalar couplings. Their beta functions are given by
\begin{align}
\beta_{u_{1}} & \,\,=\,\,4\alpha_{u_{1}}\left[8\alpha_{u_{1}}+8\alpha_{u_{2}}+\left(\alpha_{y}+\alpha_{\tilde{y}}\right)\right]+32\alpha_{v_{1}}^{2}x_{F}^{2}+6\alpha_{u_{2}}^{2}\,\,,\nonumber \\
\beta_{u_{2}} & \,\,=\,\,2\alpha_{u_{2}}\left[4\alpha_{u_{2}}+\left(\alpha_{y}+\alpha_{\tilde{y}}\right)\right]-\frac{1}{2}x_{F}(\alpha_{y}^{2}+\alpha_{\tilde{y}}^{2})\,\,,\nonumber \\
\beta_{w_{1}} & \,\,=\,\,4\alpha_{w_{1}}\left[8\alpha_{w_{1}}+24\alpha_{w_{2}}+2x_{F}(\alpha_{Y}+\alpha_{\tilde{Y}})-3\alpha_{g}\right]+32\alpha_{v_{1}}^{2}x_{F}^{2}+48\alpha_{w_{2}}^{2}+\frac{3}{8}\alpha_{g}^{2}\,\,,\nonumber \\
\beta_{w_{2}} & \,\,=\,\,2\alpha_{w_{2}}\left[12\alpha_{w_{2}}+2x_{F}(\alpha_{Y}+\alpha_{\tilde{Y}})-3\alpha_{g}\right]-\frac{1}{2}x_{F}(\alpha_{Y}^{2}+\alpha_{\tilde{Y}}^{2})+\frac{3}{16}\alpha_{g}^{2}\,\,,\nonumber \\
\beta_{v_{1}} & \,\,=\,\,2\alpha_{v_{1}}\left[16\alpha_{u_{1}}+8\alpha_{u_{2}}+16\alpha_{w_{1}}+24\alpha_{w_{2}}+(\alpha_{y}+\alpha_{\tilde{y}})+2x_{F}(\alpha_{Y}+\alpha_{\tilde{Y}})-3\alpha_{g}\right]
\nonumber \\ &\qquad\qquad\qquad\qquad\qquad\qquad\qquad\qquad\qquad\qquad\qquad
-\frac{1}{2}(\alpha_{y}+\alpha_{\tilde{y}})(\alpha_{Y}+\alpha_{\tilde{Y}})-\Upsilon\,\,.
\end{align}
Analysis of these RGEs shows that there is only a real solution for
a fixed point when $\alpha_{Y}=\alpha_{\tilde{Y}}=0$, corresponding
to fixed point B, namely the original LS fixed point studied in \cite{Abel:2017rwl}.
Along the trajectory from B, the couplings assume the following values
(with actually two stable branches for $w_{1}$):
\begin{align}
\alpha_{u_{1}} & ~=~\frac{-6\sqrt{22}+3\sqrt{19+6\sqrt{22}}}{100}\alpha_{g}\,\,,\nonumber \\
\alpha_{u_{2}} & ~=~\frac{3}{25}\left(\sqrt{22}-1\right)\alpha_{g}\,\,,\nonumber \\
\alpha_{w_{1}} & ~=~\frac{3\pm\sqrt{3\left(4\sqrt{2}-5\right)}}{16\sqrt{2}}\alpha_{g}\,\,,\nonumber \\
\alpha_{w_{2}} & ~=~\frac{1}{16}\left(2-\sqrt{2}\right)\alpha_{g}\,\,, \nonumber \\ 
\alpha_{v_{1}} & ~=~0\,\ .\label{eq:quartics}
\end{align}
It is important for later reference that, as discussed in  \cite{Litim:2014uca,Litim:2015iea}, the 
negative value of $\alpha_{u_{1}}$ at the minimum does not induce instability in $H$ because it is off-set by the much larger positive value of $\alpha_{u_{2}}$.

\begin{figure}
\centering{}\includegraphics[scale=0.5]{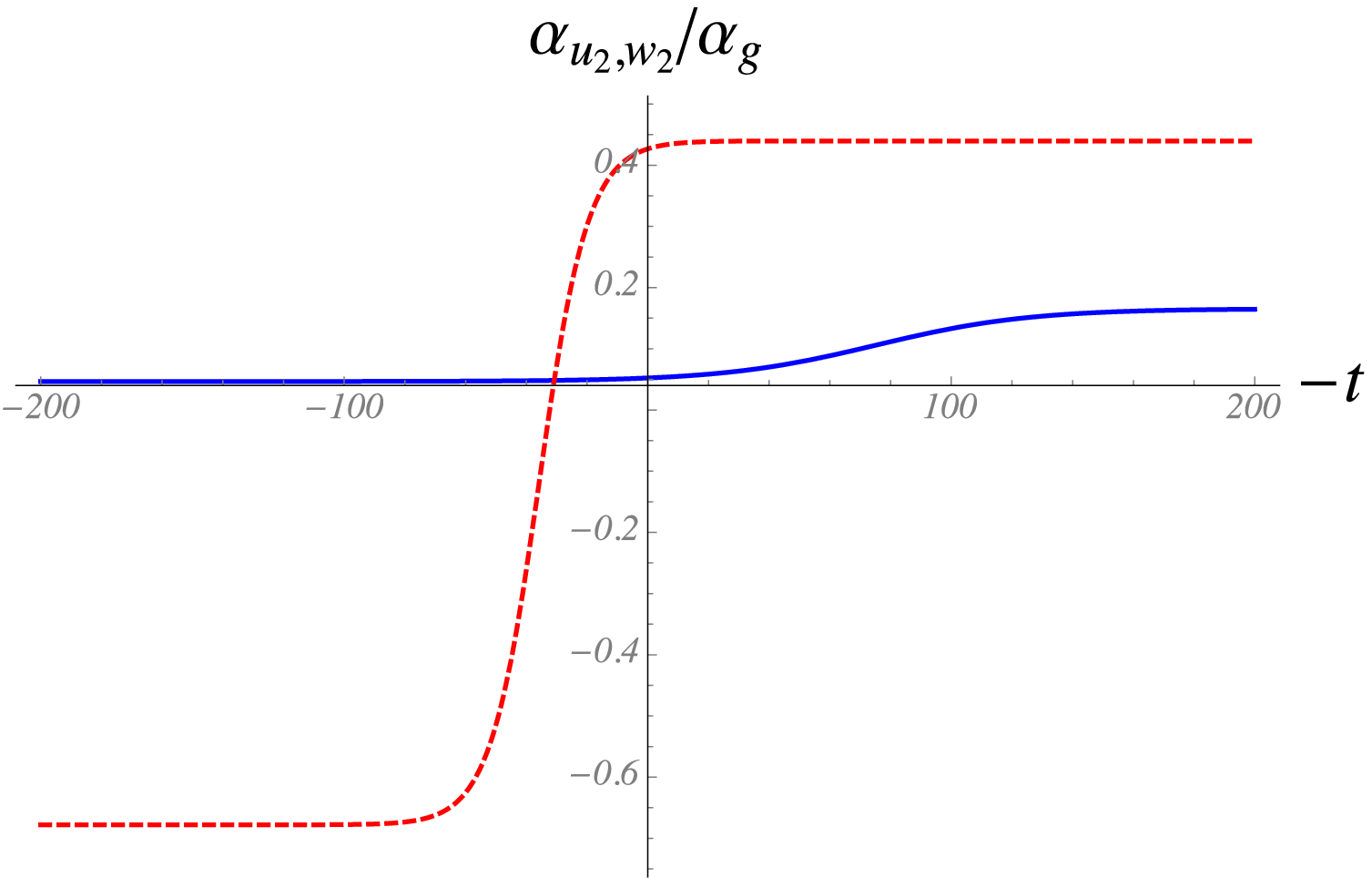}
\includegraphics[scale=0.5]{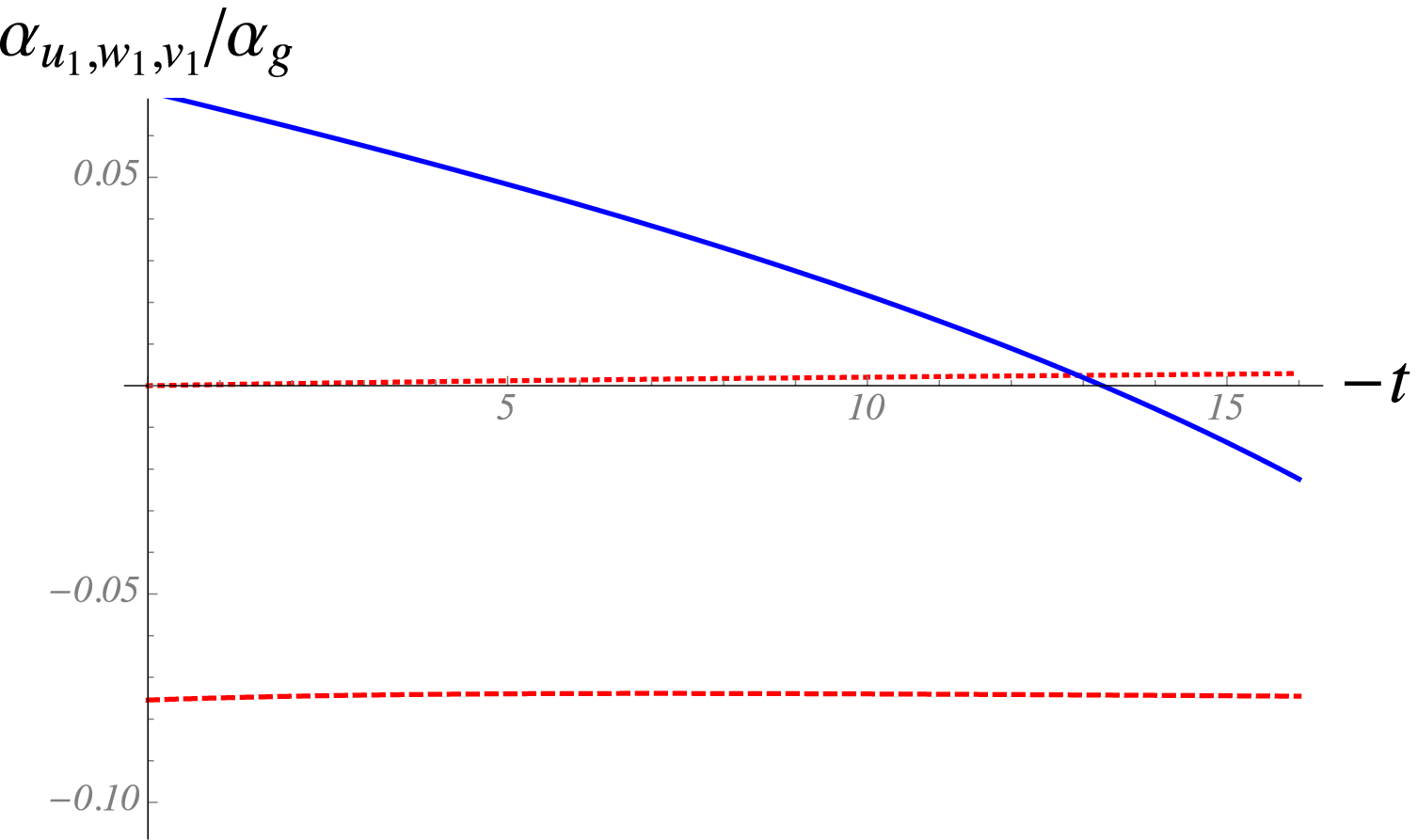}
\caption{\emph{The 
flow for the quartic couplings once the theory leaves trajectory B (at $t=0$). On the left, $\alpha_{u_2}$ in red/dashed and $\alpha_{w_2}$ in blue/solid both flow to new fixed points (beginning from positive values at $t=0$). The $\alpha_{w_2}$ quasi-fixed point is much larger, which among other contributions induces $\alpha_{w_1}$ in blue/solid (on the right) to run negative and form a minimum radiatively for $\tilde{S}$. Meanwhile  $\alpha_{u_1}$ in red/dashed on the right is only moderately changed, not enough to destabilise $H$, while $\alpha_{v_1}$ in red/dotted runs slightly positive, inducing a small positive mass-squared for the Higgs at the PS breaking scale.  This example has $\epsilon=0.01$.
 \label{fig:quarticflow2}}}
\end{figure}

\begin{figure}
\centering{}\includegraphics[scale=0.5]{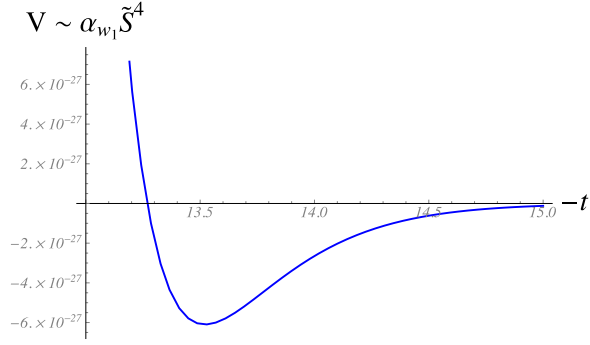}
\caption{\emph{The  radiatively induced minimum in the effective potential for $\tilde{S}$.
 \label{fig:quarticflow3}}}
\end{figure}

Once the Yukawa flow in Figure~\ref{fig:The-flow-in} begins, the scalar couplings also begin to flow: indeed Figure~\ref{fig:The-flow-in} is somewhat idealised in the sense that the quartic couplings now rapidly induce radiative symmetry breaking.
In order to show this analytically, one may use the relations between the Yukawas
and gauge couplings corresponding to trajectory E which yields an effective
set of beta functions:
\begin{align}
\label{betaff}
\beta_{u_{1}} & \,\,=\,\,32\alpha_{u_{1}}^{2}+6\alpha_{u_{2}}^{2}+32\alpha_{u_{1}}\alpha_{u_{2}}+\frac{1056}{277}\alpha_{u_{1}}\alpha_{g}\,,\nonumber \\
\beta_{u_{2}} & \,\,=\,\,8\alpha_{u_{2}}^{2}+\frac{528}{277}\alpha_{u_{2}}\alpha_{g}-\frac{182952}{76729}\alpha_{g}^{2}\,,\nonumber \\
\beta_{w_{1}} & \,\,=\,\,32\alpha_{w_{1}}^{2}+48\alpha_{w_{2}}^{2}+96\alpha_{w_{1}}\alpha_{w_{2}}-\frac{2820}{277}\alpha_{w_{1}}\alpha_{g}+\frac{3}{8}\alpha_{g}^{2}\,,\nonumber \\
\beta_{w_{2}} & \,\,=\,\,24\alpha_{w_{2}}^{2}-\frac{1410}{277}\alpha_{w_{2}}\alpha_{g}+\frac{227163}{1227664}\alpha_{g}^{2}\, , \nonumber \\
\beta_{v_{1}} & \,\,=\,\,16\ \alpha_{v_{1}} 
\left( 2\alpha_{u_{1}}+ \alpha_{u_{2}}+2\alpha_{w_{1}}+ 3 \alpha_{w_{2}} -\frac{441}{2216}\alpha_{g}\right) 
-\frac{1584}{76729}\alpha_{g}^{2}
\,.
\end{align}
These show that, once the system is kicked onto the E trajectory, the quartic couplings $u_{2}$ and $w_{2}$ flow to 
 ``quasi-fixed points'', that is trajectories that are determined entirely by the slowly varying value of $\alpha_g$. 
Indeed $\alpha_g$ is parametrically slowly flowing compared to the quartics (because its beta function is order $\epsilon^2$), so we may approximate it as  
constant, with the quartic couplings starting close to the boundary values  in (\ref{eq:quartics}).
Solving for 
$\alpha_{u_2}$ and $\alpha_{w_2}$ we see that they can both asymptote (as $\tanh$ functions) 
to positive ``quasi-fixed'' IR values given by\footnote{A more sophisticated treatment is possible, and other flows are possible,  but this prescription is sufficient for a qualitative understanding.} 
\begin{align}
\frac{\alpha_{u_2}}{\alpha_g} &~\approx~  - \frac{33}{277} \left( 1+\sqrt{22} \tanh\left( \frac{264\sqrt{22}}{277}  \frac{\Delta t}{\alpha_g} \right) \right)
~\qquad\qquad\longrightarrow ~~  \frac{33}{277} ( \sqrt{22} -1) ~, \nonumber \\
\frac{\alpha_{w_2}}{\alpha_g} &~\approx ~  \frac{1}{4432} \left( 470- \sqrt{69458} \tanh\left( \frac{3\sqrt{34729}}{277\sqrt{2}}  \frac{\Delta t}{\alpha_g} \right)\right)~~
~~~\longrightarrow ~~  \frac{1}{4432} ( \sqrt{69458} +470) ~ .  
\end{align}
While the first quasi-fixed value for $u_2$ is very close to its starting point in \eqref{eq:quartics}, the quasi-fixed value for $w_2$ is much larger. 
(We should repeat that the analysis is approximate because  the Yukawa couplings have not yet reached their new trajectory; in a full numerical evolution the running of the quartics will be delayed because the Yukawas have not yet reached trajectory E, but it is expected to be qualitatively the same.) This evolution is shown numerically in the left panel  of 
Figure~\ref{fig:quarticflow2}. In the right panel we show the effect on the remaining couplings $u_1,w_1,v_1$. Because $w_2$ appears only in the RGE for $w_1$, the two couplings $u_1$ and $v_1$ are changed only very slightly, with $v_1$ becoming slightly positive. Importantly no instability can be induced radiatively for $H$ at this stage, because (as was the case above on trajectory B) the negative contribution of $u_1$ to the potential is still off-set by the positive contribution of $u_2$. On the other hand $w_2$ runs more positive in the IR,
and as can be seen from \eqref{betaff}, this is a positive contribution to $\beta_{w_1}$, adding to several other positive contribution to $\beta_{w_1}$. The nett result is that  $w_1$ runs negative (regardless of $w_2$ in fact) and inevitably at some point overcomes the positive approximately constant contribution to the potential from the $w_2$ term itself, forming a radiative minimum as in Figure~\ref{fig:quarticflow3} (as per \cite{gildener2}). 

Thus  (extended) PS breaking is induced radiatively in the TM, and at this scale a small positive mass-squared is generated via the $v_1$ ``portal'' coupling. It is natural for the latter to then be driven negative itself  below the PS breaking scale, due to the coupling of $H$ to the $N_F-n_gN_S \approx \frac{9}{4} N_C$ pairs of  $Q$,$\tilde{Q}$ fields that remain light because (by chiral symmetry) they cannot receive a mass from the $Y$,$\tilde{Y}$ couplings. This flow would be similar to that of the other mass-squared operators above the scale of PS breaking which we discuss in the following subsection: a more complete analysis of the running of the ``portal'' Higgs mass-squared  below the PS scale will be undertaken in a later phenomenological study.  

\subsection{\it The behaviour of the relevant couplings -- emergent flavour hierarchies from flavour
symmetric fixed points}

We now turn to the behaviour of classically relevant operators. These are allowed in the 
model as they are unable to disrupt the UV fixed point. (In any asymptotically safe theory such classically relevant 
operators are simply part of the collection of non-predictive parameters in the theory.) They renormalise multiplicatively
and can themselves initiate radiative symmetry breaking, as described in  \cite{Abel:2017ujy}. 
As in the minimal supersymmetric SM,  one can begin with a set of entirely positive mass-squareds
in the UV and have them run negative due to the large Yukawa couplings in the model.

In this subsection we shall perform a more complete analysis of the flow of these operators to show how one should  incorporate their flavour dependence. 
In particular we are interested in the possible generation of flavour/generation hierarchies in $H$, which in any viable model will be required to satisfy 
phenomenological constraints.

 Rather than write the explicit flavour dependence as in (\ref{eq:lsoft}), we wish to 
consider smaller flavour structures that are closed under
RG. To see how to do this, as a warm-up consider the completely $SU(N_F)$ symmetric terms in \cite{Abel:2017rwl}, which were mass-squareds
of the form 
\begin{align}
M_{H_{ij}H^*_{kl}}^{2}\, & =\,m_{0}^{2}\delta_{jl}\delta_{ik}+2\Delta^{2}\sum_{a}T_{ji}^{a}T_{kl}^{a}\nonumber \\
 & =\,m_{0}^{2}\delta_{jl}\delta_{ik}+\Delta^{2}\left(\delta_{jl}\delta_{ik}-\frac{1}{N_{F}}\delta_{ji}\delta_{kl}\right)~.\label{eq:masssquared}
\end{align}
Defining real and imaginary parts, $H_{ij}=\frac{1}{\sqrt{2}}\left(h_{ij}+ip_{ij}\right)$, and\footnote{Note that $h_{a}$ and $p_{a}$ are not simply related to $h_{ij}$
and $p_{ij}$. That is, while $p_{a}$ is the coefficient of the
anti-hermitian parts of $H$, $p_{ij}$ is the coefficient of the
\emph{imaginary} parts of $H$}  $h_{a}+ip_{a}=\sqrt{2}T_{ij}^{a}(h_{ij}+ip_{ij})$, the corresponding operators can be written as
\begin{align}
m_{0}^{2}H^{\dagger}H & \,=\,\frac{m_{0}^{2}}{2}\text{Tr}\left(h^{2}+p^{2}\right)~,\nonumber \\
\sum_{a}\frac{\Delta^{2}}{2}\left(h_{a}^{2}+p_{a}^{2}\right)\,=\,\sum_{a}2\Delta^{2}\text{Tr}(T_{a}H)\text{Tr}(T_{a}H^{\dagger}) & \,=\,\frac{\Delta^{2}}{2}\left[\text{Tr}\left(h^{2}+p^{2}\right)-\frac{\left(\text{Tr}h\right)^{2}+\left(\text{Tr}p\right)^{2}}{N_{F}}\right]\,.
\label{oldie}
\end{align}

It is now clear that one can proceed to break flavour in a way that commutes with the RG equations, by arranging the flavour breaking 
in $SU(n)$ subgroups, where the $SU(n)$ generators are in the $n\times n$ upper-left $n\times n$ block of the parent $SU(N_F)$,  where $1<n\leq N_{F}$.   
This gives degenerate masses for the generators of each nested $SU(n)$ flavour subgroup,
where we envisage an explicit breaking 
\[
SU(N_{F})\supset SU(N_{F}-1)\ldots\supset SU(n)\ldots
\]
So without loss of generality we can express the new Cartan generators introduced
for each $SU(n)$ as 
\begin{equation}
T_{ij}^{(n^{2}-1)}=\frac{1}{\sqrt{2n(n-1)}}\left(\begin{array}{cccccc}
1\\
 & \ddots\\
 &  & 1\\
 &  &  & 1-n\\
 &  &  &  & 0\\
 &  &  &  &  & \ddots
\end{array}\right)\,,
\end{equation}
with the non-Cartan generators being defined accordingly in the obvious way. 
Defining the trace over the $SU(n)$ block of the generators as 
\begin{equation}
\text{Tr}_{n}(\mathcal{O}_{ij})=\sum_{i=1}^{n}\mathcal{O}_{ii}\,,
\end{equation}
the flavour breaking generalisation of \eqref{oldie} becomes 
\begin{align}
V^{(2)} &~ = ~\,\frac{m_{0}^{2}}{2}\text{Tr}_{N_{F}}\left(h^{2}+p^{2}\right)+\sum_{n=1}^{N_{F}-1}\frac{m_{n}^{2}}{2}\left[\frac{\left(\text{Tr}_{n}h\right)^{2}+\left(\text{Tr}_{n}p\right)^{2}}{n}\right]\nonumber \\
& \qquad\qquad\qquad\qquad\qquad\qquad\qquad\qquad\qquad +~\sum_{n=2}^{N_{F}}\frac{\Delta_{n}^{2}}{2}\left[\text{Tr}_{n}\left(h^{2}+p^{2}\right)-\frac{\left(\text{Tr}_{n}h\right)^{2}+\left(\text{Tr}_{n}p\right)^{2}}{n}\right]\,.\label{eq:vclass-master}
\end{align}
\begin{center}
\begin{table}
\begin{centering}
\caption{\it The relevant quadratic  operators and their beta function coefficients in terms of the quartic $u_{1,2}$ couplings.\label{tab:Relevant-operators-and}}
\begin{tabular}{|c|c|c|}
\hline 
{\scriptsize{}coupl'g} & {\scriptsize{}Operator} & {\scriptsize{}Coefficient in $16\pi^{2}\partial_{t}V$}\tabularnewline
\hline 
{\scriptsize{}$m_{0}^{2}$} & {\scriptsize{}$Tr_{N_{F}}(h^{2}+p^{2})$} & {\scriptsize{}$\begin{array}{c}
m_{0}^{2}\left\{ 2u_{1}\left[N_{F}^{2}+1\right]+4u_{2}N_{F}\right\} +\Delta_{N_{F}}^{2}\left(2u_{1}+\frac{4u_{2}}{N_{F}}\right)(N_{F}^{2}-1)\,\,\,\,\,\,\,\,\,\,\,\,\\
\,\,\,\,\,\,\,\,\,\,\,\,\,\,\,\,\,\,+\,\sum_{n}^{N_{F}-1}2u_{1}\left(m_{n}^{2}+\Delta_{n}^{2}\left(n^{2}-1\right)\right)
\end{array}$}\tabularnewline
\hline 
{\scriptsize{}$\Delta_{N_{F}}^{2}$} & {\scriptsize{}$\text{Tr}_{N_{F}}\left(h^{2}+p^{2}\right)-\frac{\left(\text{Tr}_{N_{F}}h\right)^{2}+\left(\text{Tr}_{N_{F}}p\right)^{2}}{N_{F}}$} & {\scriptsize{}$2u_{1}\Delta_{N_{F}}^{2}$}\tabularnewline
\hline 
{\scriptsize{}$\Delta_{n}^{2}$} & {\scriptsize{}$\text{Tr}_{n}\left(h^{2}+p^{2}\right)-\frac{\left(\text{Tr}_{n}h\right)^{2}+\left(\text{Tr}_{n}p\right)^{2}}{n}$} & $2u_{1}\Delta_{n}^{2}+\frac{4u_{2}}{n}\left(m_{n}^{2}+\Delta_{n}^{2}\left(n^{2}-1\right)\right)$\tabularnewline
\hline 
{\scriptsize{}$m_{n}^{2}$} & $\frac{\left(\text{Tr}_{n}h\right)^{2}+\left(\text{Tr}_{n}p\right)^{2}}{n}$ & $2u_{1}m_{n}^{2}+\frac{4u_{2}}{n}\left(m_{n}^{2}+\Delta_{n}^{2}\left(n^{2}-1\right)\right)$\tabularnewline
\hline 
\end{tabular}
\par\end{centering}{\scriptsize \par}
\end{table}
\end{center}
These operators form a system closed under RG, and we may now determine their coefficients 
in $16\pi^{2}\partial_{t}V$, relevant for solving the Callan-Symanzik equation: these are shown in Table \ref{tab:Relevant-operators-and}. 
One can now solve the RG equations along trajectory B for these parameters
to see how they evolve before their flow is cut off by the radiative symmetry breaking (regardless of how it arises): as for any relevant parameter the
flow will be expressed in terms of a set of RG-invariants. In this case, defining
 $f_{y}=\alpha_{y}/\alpha_{g}\approx0.46$, $f_{u_{1}}=\alpha_{u_{1}}/\alpha_{g}\approx-0.30$,
$f_{u_{2}}=\alpha_{u_{2}}/\alpha_{g}\approx0.44$, and 
\begin{align}
f & \,\,=\,\,2f_{y}+4f_{u_{1}}\left(1+\frac{1}{N_{F}^{2}}\right)+8f_{u_{2}}\,\approx\,3.22\,\, , \nonumber \\
f_{\Delta} & \,\,=\,\,2f_{y}+\frac{4}{N_{F}^{2}}f_{u_{1}}\,\approx\,0.92~, \nonumber \\
f_{n} & \,\,=\,\,8f_{u_{2}}\frac{n}{N_{F}}(1-\delta_{nN_{F}})\,\,,
\end{align}
the RG-invariants are found to be
\begin{align}
\tilde{m}_{*}^{2} & =\tilde{m}^{2}(0)\left(\tilde{\Omega}(0)\right)^{-f},\\
\sigma_{n*}^{2} & =\left[m_{n}^{2}(0)+(n^{2}-1)\Delta_{n}^{2}(0)\right]\left(\tilde{\Omega}(0)\right)^{-(f_{\Delta}+f_{n})},\\
\rho_{n*}^{2} & =\left[\Delta_{n}^{2}(0)-m_{n}^{2}(0)\right]\left(\tilde{\Omega}(0)\right)^{-f_{\Delta}},
\end{align}
where 
\begin{equation}
\tilde{\Omega}(t)\,=\,\left(\frac{\alpha_{g}^{*}}{\alpha_{g}}-1\right)^{-3/4\epsilon}.
\end{equation}
In terms of these we find the following solutions for the operators in \eqref{eq:vclass-master}: 
\begin{align}
m_{0}^{2} & =\left(\frac{\tilde{\Omega}(t)}{\tilde{\Omega}(0)}\right)^{f}\tilde{m}_{*}^{2}-\frac{1}{N_{F}^{2}}\sum_{n}^{N_{F}}\frac{\sigma_{n*}^{2}}{1+2\frac{f_{u_{2}}}{f_{u_{1}}}(1-n/N_{F})}\tilde{\Omega}^{f_{\Delta}+f_{n}}\,,\nonumber \\
\Delta_{n}^{2} & =\frac{1}{n^{2}}\left(\rho_{n*}^{2}\tilde{\Omega}^{f_{\Delta}}+\sigma_{n*}^{2}\tilde{\Omega}^{f_{\Delta}+f_{n}}\right)\,,\nonumber \\
m_{n}^{2} & =\frac{1}{n^{2}}\left(\rho_{n*}^{2}(1-n^{2})\tilde{\Omega}^{f_{\Delta}}+\sigma_{n*}^{2}\tilde{\Omega}^{f_{\Delta}+f_{n}}\right)\,.
\end{align}
This is the desired form,
since it assumes nothing about the ``starting values'', which are
simply values chosen at an arbitrary point in renormalization time, and it properly encapsulates all the non-predictive parameters in the theory. The 
entire flow is determined by these parameters and $\tilde{\Omega}(t)$, which just determines where one is in renormalization time.

The interesting feature of these solutions is that $\tilde{\Omega}\rightarrow0$ in the IR. Simple flavour hierarchies
can therefore be generated much like the mechanism for radiative symmetry breaking
in \cite{Abel:2017ujy}. That is the exponent $f$ is much larger than $f_{n}$
or $f_{\Delta}$. Therefore $m_{0}^{2}$ runs to zero in the IR
much more quickly than $\Delta_{n}^{2}$ or $m_{n}^{2}$. Meanwhile
in the deep IR one can see from these solutions and the corresponding
operators in Table \ref{tab:Relevant-operators-and}, that hierarchies are 
naturally driven into the trace components in the potential which becomes dominant:
\begin{equation}
V~\longrightarrow~\sum_{n>1}\Delta_{n}^{2}\left[\text{Tr}_{n}\left(h^{2}+p^{2}\right)-n\left(\left(\text{Tr}_{n}h\right)^{2}+\left(\text{Tr}_{n}p\right)^{2}\right)\right]~.
\end{equation}
This supports the intriguing possibility that flavour hierarchies originate within the VEVs of the Higgs sector, which would themselves become correspondingly hierarchical. 

\section{Conclusions}

In this paper we have presented a model, the Tetrad Model (TM), which is asymptotically safe, and which descends directly to the Standard Model via radiative symmetry breaking. In terms of a convenient ``quiver-like'' interpretation, the model contains 4-units, with matter and electroweak Higgs fields falling into an extended Pati-Salam GUT structure, based on the gauge group $SU(N_C)\times SU(2)_L\times SU(2)_R$, and a fourth unit that provides the PS breaking. (The electroweak gauging of a subgroup of the flavour symmetry can not be shown on the quiver, but nevertheless the language is useful for understanding the overall structure of the gauge-Yukawa UV fixed point.) At low energies the model is able to yield the Standard Model enhanced only in the Higgs fields, which carry the same generation indices as the matter fields.

Remarkably radiative symmetry breaking (i.e. the Coleman-Weinberg mechanism) operates in the model with no further adjustment. The (extended) PS Higgs naturally develops a VEV radiatively while the electroweak Higgs gains a positive ``boundary value'' mass-squared at the PS scale, due to a portal coupling that runs from zero at the UV fixed point. This mass-squared can itself then be driven negative in the IR. It was also found that it is natural to generate hierarchies among the electroweak Higgs VEVs due to the enhancement of mass-squared hierarchies as the theory runs to the IR.     

\subsection*{Acknowledgements}It is a pleasure to thank Giacomo Cacciapaglia, Daniel Litim and Zhi Wei Wang for discussions. This work is partially supported by the Danish National Research Foundation under grant DNRF:90.

\end{document}